\documentstyle[11pt,epsf,epsfig]{article}
\def\gappeq{\mathrel{\rlap {\raise.5ex\hbox{$>$}}
{\lower.5ex\hbox{$\sim$}}}}

\def\lappeq{\mathrel{\rlap{\raise.5ex\hbox{$<$}}
{\lower.5ex\hbox{$\sim$}}}}

\def\ga{\mathrel{\raise.3ex\hbox{$>$\kern-.75em\lower1ex\hbox{$\sim$}}}}
\def\la{\mathrel{\raise.3ex\hbox{$<$\kern-.75em\lower1ex\hbox{$\sim$}}}}
\def\gev{{\rm \, Ge\kern-0.125em V}}
\def\tev{{\rm \, Te\kern-0.125em V}}
\def\beq{\begin{equation}}
\def\eeq{\end{equation}}
\def\beqar{\begin{eqnarray}}
\def\eeqar{\end{eqnarray}}

\def\md{M_{D}}

\newcommand\iso[2]{\mbox{${}^{#2}${\rm #1}}}
\newcommand\he[1]{\iso{He}{#1}}
\newcommand\li[1]{\iso{Li}{#1}}

\newcommand\pref[1]{(\ref{#1})}

\newcommand\msol{\hbox{$M_{\odot}$}}

\newcommand\EE[2]{{#1}\!\times\! 10^{#2}}

\begin{document}

\begin{titlepage}
\pagestyle{empty}

\baselineskip=21pt

\rightline{astro-ph/0203240}
\vskip 0.35in

\begin{center}
{\large{\bf Constraining Strong Baryon--Dark Matter Interactions with
\\ Primordial Nucleosynthesis and Cosmic Rays}}
\end{center}

\begin{center}
\vskip 0.05in
{{\bf Richard H. Cyburt}$^1$, 
{\bf Brian D. Fields}$^{2}$,
{\bf Vasiliki Pavlidou}$^2$, 
{\bf Benjamin D. Wandelt}$^{1,2,3}$ 

\vskip 0.05in
{\it
$^1${Department of Physics\\ University of Illinois, Urbana, IL 61801,
USA}
\\
$^2${Department of Astronomy\\ University of Illinois, Urbana, IL 61801,
USA}
\\
$^3${Department of Physics \\ Princeton University, Princeton, NJ 08544}}}

\vskip 0.35in
{\bf Abstract}
\end{center}
\baselineskip=18pt \noindent

Self-interacting dark matter (SIDM) was introduced by Spergel \& Steinhardt
to address possible discrepancies between
collisionless dark matter simulations and observations on scales of
less than 1 Mpc.  We 
examine the case in which dark matter particles not only have strong
self-interactions but also have strong interactions with baryons.  The
presence of such interactions will have direct implications for
nuclear and particle astrophysics.  Among these are a change in the
predicted abundances from big bang nucleosynthesis (BBN) and the flux
of $\gamma$-rays produced by the decay of neutral pions which originate
in collisions between dark matter and Galactic cosmic rays (CR).  From these
effects we constrain the strength of the baryon--dark matter
interactions through the ratio of baryon - dark matter interaction
cross  section to dark matter mass, $s$.  We find that BBN places a weak upper
limit to this ratio $\la 10^8 {\rm \, cm^2 \, g^{-1}}$.  CR-SIDM
interactions, however, limit the possible DM-baryon cross section to
$\la 5 \times 10^{-3}  {\rm \, cm^2 \, g^{-1}}$;
this rules out an energy-independent interaction,
but not one which falls with center-of-mass velocity
as $s \propto 1/v$ or steeper.

\noindent
{\em PACS Numbers:}
26.35.+c; 
95.30.Cq; 
95.35.+d; 
98.70.Sa; 
13.85.Tp  

\vfill

\end{titlepage}
\baselineskip=18pt

\section{Introduction}

Compelling observational evidence suggest that matter in the universe
is dominated by ``cold dark matter'' (CDM), the simplest model of
which is one where dark matter particles have 
interaction strengths at or below the weak scale, and
thus interact today only through gravity ({\em collisionless} dark
matter).  The collisionless cold dark matter model has been proven
through analytic and numeric simulations to be very successful in
explaining  the large scale structure of the universe
\cite{OstrikerSteinhardt}.

However, a series of observations of dark matter structures on the order of
$\la 1$ Mpc (e.g., galaxy rotation curves \cite{flores,dalc,deblok},
strong lensing from galaxy clusters \cite{tyson}, Tully-Fisher
relation of spiral galaxies \cite{mo}, bar stability in spirals
\cite{deb1,deb2} and deficit of low-mass subhalos in the Local Group
\cite{klypin}) appear to contradict the prediction of collisionless
CDM simulations that dark matter halos form high-density cores,
favoring shallower, lower-density dark matter halo  cores.

Spergel and Steinhardt \cite{SS} have proposed a variation of CDM in
order to alleviate these discrepancies. In their proposed picture,
dark matter particles interact with each other strongly and have a
large scattering cross section, but negligible annihilation or
dissipation.\footnote{
If the dark matter particle and antiparticle states were equally
populated and annihilations were allowed, these
would proceed vigorously, leaving a relic abundance today
$\Omega_{\rm D} \sim 
  10^{-37} \ {\rm cm^2} \langle \sigma_{\rm ann} v/c \rangle^{-1}$ 
which would be negligible for the case of strong interactions.
This implies that the SIDM particles either have
strongly suppressed annihilation cross sections, or 
are asymmetrically populated (as is the case for, e.g.,
baryons).
}
Subsequent numerical simulations \cite{dave,wand} have
shown that this collisional cold dark matter, or SIDM,
does indeed predict halo cores in better agreement with the
observations mentioned above, provided 
\beq
\label{eq:s-halo}
0.5 < s < 6 {\rm \, cm^2 \, g^{-1}}. 
\eeq
where the
the parameter $s = \sigma_{\rm
SIDM} / \md$ is
the dark matter--dark matter
elastic scattering cross section 
over the dark matter particle mass.
Potential difficulties for this model may
arise at cluster scales \cite{arcs,tyson,yswt,wts};
one way to address these would be a velocity-dependent
cross section, the effect of which we will consider below.

If dark matter particles interact with each other through the strong
force, then similar interactions would be expected between them and
ordinary baryons as well, with cross sections of the same order
\cite{wand}.  Wandelt et al.\ \cite{wand} considered
this case and found, perhaps surprisingly, that such interactions
cannot be excluded by galaxy halo data nor from {\em direct} observation by
space-borne cosmic-ray detectors, if dark matter particles have a mass
larger than $10^5$ GeV.  Earlier papers \cite{sged} have also
considered astrophysical constraints on SIDM both with and without
interactions with baryons. 

{}From the particle physics point of view, we choose a model-independent
approach. The question we address is ``If a massive neutral particle
exists, how can we constrain its properties from astrophysical
observations?''.  We refrain from a detailed discussion of the naturalness
of such a particle except to note how quickly model builders proposed
several particle physics implementations (e.g. \cite{sidm}) of
self-interacting dark matter. 
Some promising candidates, the $S0$ and low-mass strangelets,
were then ruled
out \cite{wand}. 
Nevertheless, there remain
intriguing candidates for dark
matter which lie in the regions of masses and
baryon/self-interaction cross sections we consider, both within
supersymmetric extensions of the standard model \cite{kusenko} or as
possible relics from the QCD phase transition within the standard model
\cite{zhitnitsky}.  Our results place constraints
on the energy dependences of the cross sections in these models.

In this paper, we examine the effect of inelastic SIDM--baryon interactions.
Specifically, we consider the impact of such interactions on big bang
nucleosynthesis (BBN), and on the $\gamma-$rays produced
by interactions of cosmic ray nuclei with dark matter particles.
In BBN, the formation of the light elements deuterium, \he3, \he4, and \li7,
depends on the time at which the weak interactions ``freeze out,'' the
free neutron decay rate, and the temperature at which nucleosynthesis
occurs, the latter being determined by the
photo-dissociation of deuterium.  If dark matter
can interact strongly with baryons, it is possible that it can destroy
light nuclei.  Since deuterium has the smallest binding energy and
determines the onset of nucleosynthesis, any dark matter deuterium
dissociation will change where the onset of nucleosynthesis happens
and thus provide a sensitive probe to baryon--dark matter interactions
during BBN.  We will see that the constraints provided by BBN are
quite weak--in other words, for the parameter ranges of interest, SIDM
has negligible effect on BBN despite the strength of the SIDM-baryon
interaction.  

The interaction of cosmic rays with
interstellar matter leads to observable signatures
in the form of $\gamma-$rays.
Collisions between
energetic cosmic ray protons and interstellar hydrogen can
produce neutral pions, which in turn decay into two high-energy $\gamma-$rays;
this emission is believed to dominate the observed flux
of high-energy photons from the Galactic disk
\cite{sreekumar,hunter,strong}.
Including dark matter--baryon interactions allows for similar
processes to occur between cosmic ray nuclei and ambient SIDM, thus possibly
contributing to the $\gamma-$ray diffuse background. 
The $\gamma-$ray flux from CR-SIDM interactions 
will be a function of the assumed cross section for the interaction 
between dark and baryonic matter and  
can be calculated given the functional form of the Galactic 
cosmic ray flux and dark matter distribution. The result can then be 
compared with diffuse $\gamma$-ray observations 
from $\gamma-$ray telescopes (most recent of which are the EGRET 
observations,
\cite{hunter,sreekumar})
and strong upper limits can be placed on the cross sections.

This paper is organized as follows.  We will discuss the strong
interacting dark matter cross section in \S \ref{sect:CS}.
In \S \ref{sect:BBN} we determine the SIDM effect on BBN and its
constraints.  In \S \ref{sect:CR}  we compute the expected
$\gamma$-ray flux from cosmic ray - SIDM  interaction in the Milky Way
and the resulting constraints on the  cross section. 
Other possible constraints are noted in \S \ref{sect:other}.
Finally, our
findings are summarized and discussed  in \S \ref{sect:dis}.

%
%

\section{SIDM-Nucleon Interactions}
\label{sect:CS}

We wish to characterize the inelastic interactions between
baryonic nuclei and SIDM particles $D$.
We will express the cross sections for
inelastic interactions in terms
of the elastic cross sections.
The inelastic behavior is assumed to 
obey simple scalings with the masses of the reactants \cite{SS}
and be energy-independent (but we will use the different
constraints below to infer constraints on possible energy dependence
of SIDM-baryon interactions).

We begin by considering the SIDM-nucleon
interaction, with an energy-independent 
elastic scattering cross section 
$\sigma_{DN}^{\rm elastic}$.
This need not be the same as the SIDM self-interaction
cross section, but was taken to be so in the
work of \cite{wand}.
One can then generalize this to determine the
cross section for elastic SIDM-nucleus scattering.
Following \cite{sged}, we adopt the form  
\beq
\label{eqn:CS-DA}
\sigma_{DA}^{\rm elastic} = A^2\left( \frac{\mu (A)}{\mu (p)} \right)^2
\sigma_{DN}^{\rm elastic}
\eeq
where $A$ is the atomic mass number of the interacting nuclide, and
$\mu(A)$ is the reduced nucleus--dark matter mass.
Finally, we will be interested in {\em inelastic} scattering,
i.e., when internal degrees of freedom are excited in the 
baryonic component.  We will write the (energy-independent)
inelastic scattering cross section for channel $i$ as
\beq
\sigma_{DA}^i = \alpha_i \sigma_{DA}^{\rm elastic}
\eeq
where $\alpha_i$ encodes the relative strength of the
inelastic interaction.

A typical strong
interaction has a cross section, $\sigma_{DN}^{\rm elastic} \sim 1$ barn and it
is assumed that a dark matter particle has a mass, $\md  \ga 1$
GeV.  A useful parameter given these typical values for the cross
section and mass is defined as 
\beq
s = \sigma_{DN}^{\rm elastic}/\md
\eeq
usually
quoted in cgs units as cm$^2$ g$^{-1}$.
A useful conversion factor for particle physics units
is $1 \ {\rm cm^2 \ g^{-1}} = 1.78 \ \rm{barn}/({\rm GeV}/c^2)$.

\section{Limiting SIDM with BBN}
\label{sect:BBN}


Primordial nucleosynthesis is the process by
which the light elements, consisting mainly of deuterium, \he3, \he4, and
\li7, were produced.  
Standard BBN is a one parameter theory, depending only on the
baryon-to-photon ratio, 
$\eta \equiv n_{\rm B}/n_\gamma = \EE{2.74}{-8} \Omega_{\rm B}h^2$, 
where $\Omega_{\rm B}$ is the current
baryon density relative to the critical density, $\rho_{\rm crit} =
3H_0^2/8\pi G$, and $H_0 = 100h \ {\rm km \ s^{-1}\ Mpc^{-1}}$ is the
present value of the Hubble parameter.

Before considering the effect of SIDM, it is useful to recall the
basic physics of BBN.
At the onset ($T \ga 1$ MeV; $t \la 1$ s) of the 
BBN epoch, the universe was
dominated by relativistic particle species, initially
consisting of photons, 3 species of light neutrinos, and
$e^\pm$ pairs; baryons were non-relativistic and
essentially only $n$ and $p$ with no complex nuclei.
At this stage,
all of these
particle species existed in thermal and chemical equilibrium, due to
the interaction rates between particles being much faster than the
expansion rate of the universe.  

At a temperature around an MeV, and time about
1 s, the weak interactions, coupling the neutrinos
to the other particle species and keeping the $n$ and $p$ in equilibrium,
become ineffective (interaction rates less than
$H$).  At this time the
neutrinos fully decouple from the primordial plasma and the neutron
abundance ``freezes out'', relative to the proton abundance, with the
exception of the occasional free neutron decay.
Shortly after, the electrons and positrons start to become
non-relativistic, and thus annihilate, producing photons, which
in turn heats the plasma relative to the neutrinos.  
Trace amounts of the light
nuclei, deuterium, \he3, \he4, and \li7
exist, with abundances 
held below nuclear statistical 
equilibrium levels due to 
deuterium photo-dissociation by photons in the high-energy tail of the
thermal distribution.
This ``deuterium bottleneck'' persists until the universe cools
to $T \sim 0.07$ MeV, when 
energetic photons
are sufficiently rare, at which point
deuterium rapidly grows and
then is burnt into the other light elements.


We now consider what role CDM plays in these series of events.
Depending on the origin of the CDM and its
coupling to Standard Model particles,
CDM may or may not be in thermal equilibrium
with the rest of the cosmic plasma.
In the standard CDM scenario, 
the CDM only interacts weakly with itself and with
Standard Model particles, and thus
these interactions will freeze out well
before the epoch of BBN. Then the dark matter cools rapidly due to 
Hubble expansion. Throughout BBN, the dark matter is non-relativistic and
the Universe is radiation dominated. 
So in the standard case, dark matter plays no
role at all. 

What happens in the SIDM case? The dark matter is still
non-relativistic, but now elastic collisions between baryons and dark matter
will thermalize the dark matter at early times and bring $T_{\rm D} =
T_\gamma$. Dark matter--baryon decoupling will not
occur until much later, and we note that this will potentially have
interesting effects on structure formation\footnote{The tight coupling
of the baryon and photon fluids and the coupling 
between the dark matter and the baryons result in coupling the DM to the
photons. This gives pressure to the dark matter fluid and changes the
evolution of perturbations until the dark matter decouples from the
baryons and then cools adiabatically. We conjecture that this effect
leads to a suppression 
of fluctuation power on perturbation modes which entered the horizon
before DM-baryon decoupling. 
Standard freeze-out arguments show 
that DM-baryon decoupling occurs at a redshift of $z\sim 2\times
10^4\sigma^{\rm elastic\;{-(3/2)}}_{DN}$, where $\sigma^{\rm elastic}_{DN}$ is
measured 
in barn. The characteristic mass scale of perturbations 
entering the horizon at this time is of order $10^9
M_{\odot}(\sigma^{\rm elastic}_{DN})^2$. Note that this scaling now depends just
on the cross section, not on the mass of the particle. This 
suggests that DM-baryon interactions reduce the dark matter
perturbation power on galaxy scales --- just those which
motivated the study of interacting dark matter in the first place. We
plan to study these effects in more detail in a future publication.}. However, for
our study of BBN elastic scattering is not important except for
thermalizing the dark matter.  What does
affect BBN are {\em inelastic} interactions with SIDM particles.
Specifically, such collisions can impart
internal energy to the nucleus and lead to its dissociation. 
Given that deuterium has the smallest binding energy and
its photo-dissociation determines the onset of nucleosynthesis,
we will focus on deuterium breakup; similar processes
should occur for other nuclei but are not significant due
to the higher binding energies.

We consider the reaction $D + d \rightarrow D + n + p$
and neglect the rare, three-body, inverse reaction.  In order for
this interaction to happen the center-of-mass kinetic energy of
the dark matter--deuteron system must be equivalent to at least the
binding energy of the deuteron, $B = 2.224$ MeV.  We also assume that
the break-up cross section is comparable to the scattering cross
section, $\sigma_{Dd}^{\rm bkp} = \alpha_1\sigma_{Dd}^{\rm
elastic}$, where 
$\alpha_1$ represents the efficiency of the break-up reaction and is
of order unity.  We
relate the deuterium--dark matter scattering cross section to the
proton--dark matter cross section as defined by 
eq. (\ref{eqn:CS-DA} ).

In making this interaction applicable for BBN, we need to compute the
thermally averaged reaction rate, $\langle \sigma v\rangle$, where $v$
is the relative velocity of the interacting particles.  Since
for our discussion, we consider a constant cross section with the appropriate
energy threshold, this thermal average can be easily found.

\beq
\langle\sigma v\rangle = \EE{5.221}{9}\alpha_1\sigma_{DN}^{\rm elastic} 
\frac{\left( 1 + \frac{m_p}{\md }
\right)^2 \ }{\ \left( 1 + \frac{m_d}{\md }
\right)^{3/2}} \left( 1 + \frac{B_9}{T_9}\right)T_9^{1/2}\exp{\left(
-\frac{B_9}{T_9} \right)}\ {\rm cm^3 s^{-1}}
\eeq
where $m_p$ and $m_d$ are the masses of the proton and deuteron
respectively, and $B_9$ and $T_9$ are the binding energy of deuterium
and the temperature expressed in units of $10^9$ K.\footnote{
Throughout this analysis we assume that the dark matter remains
non-relativistic during the epoch of BBN, $\md  \ga 10$ MeV.  If
dark matter is relativistic during this time, not only will the
reaction rate take on a new form, but now the dark matter will
contribute to the expansion of the universe.}

The addition of this dark matter reaction
adds another channel for deuterium destruction in addition to
photo-disintegration.
Thus, it will take longer to build up  enough
deuterium so that nucleosynthesis can occur.  To study the effects of
the new interaction analytically, 
we will look at the quasi-static equilibrium (QSE)
abundance of deuterium, as laid out by, e.g., \cite{bbf,bbnwoc}, 
and see how this
changes when nucleosynthesis occurs.  The abundance of deuterium will
be most sensitive to the $np \rightarrow d\gamma$, 
$d\gamma \rightarrow np$ and $dD \rightarrow Dnp$ reactions.  The QSE
abundance is then given by the ratio of sources to sinks.
\beq
Y_i^{\rm QSE} \equiv n_i/n_{\rm B} = \frac{\sum_{jkl} Y_jY_k[ jk\rightarrow
il] }{\sum_{jkl} Y_l[ il\rightarrow jk]} \Rightarrow Y_d =
\frac{Y_nY_p[np\rightarrow d\gamma ]}{Y_\gamma [d\gamma\rightarrow np]
+ Y_D[dD\rightarrow Dnp]}
\eeq
Here $n_i$ is the $i^{\rm th}$ particle species number density,
$n_{\rm B}$ is the baryon number density, $Y_\gamma = 1/\eta$,
and $[jk\rightarrow il]$ are
the reaction rates 
\beq 
[jk\rightarrow il] = N_A \rho_{\rm B}\langle\sigma_{jk\rightarrow il} \,
v\rangle 
\eeq
given by several compilations, where $N_A = m_u^{-1}$ is Avogadro's
constant, and $\rho_{\rm B}$ is the baryon mass density.

The forward and reverse reaction rates are related to each other by
detailed balance.  This tells
us that the ratio of reverse to forward rates is given by the
thermal equilibrium distributions of the interacting species.  For
$np\Longleftrightarrow \gamma d$ one finds the Saha expression
\beq
\label{eq:Saha}
\frac{[d\gamma\rightarrow np]}{[np\rightarrow d\gamma ]} =
\frac{Y_n^{\rm EQ}Y_p^{\rm EQ}}{Y_d^{\rm EQ}Y_\gamma^{\rm EQ}} =
\EE{1.40}{5}T_9^{-3/2}\exp{\left( -B_9/T_9 \right) }
\eeq
Assuming dark matter has a mass $ \md \gg m_p$,
the rate is given by the following expression.
\beq 
\label{eqn:dDrate}
Y_{D}[dD\rightarrow Dnp] =
\EE{5.221}{9} \alpha_1\!
\left(\frac{s}{\rm cm^2\ g^{-1}}\right)\!\left(\frac{\Omega_{\rm
D}}{\Omega_{\rm B}}\right)\!\left(\frac{\rho_{\rm B}}{\rm g
\ cm^{-3}}\right)\! \left( 1 +
\frac{B_9}{T_9}\right)T_9^{1/2}\exp{\left( -\frac{B_9}{T_9} \right)}
\ {\rm s}^{-1}
\eeq

The ratio of the two deuterium destruction rates is then
\beq
\label{eq:d-dest}
\frac{Y_{D}[dD\rightarrow Dnp]}{Y_\gamma [d\gamma\rightarrow np]} \approx
\EE{2.1}{-9}\eta_{10}\alpha_1\!\left(\frac{s}{\rm cm^{2}\ g^{-1}}\right)\! \left(
\frac{\Omega_{\rm D}}{\Omega_{\rm B}}\right) T_9
\eeq
assuming $T_9 \ll B_9$ and where $\eta_{10} = \eta/10^{-10}$.
For reasonable values of $\alpha_1$ and $s$, we see that this
ratio is tiny for the temperatures relevant to BBN.
This simply reflects the smallness of $\eta$, i.e., the
large photon-to-baryon (and photon-to-SIDM) ratio.
The smallness of the SIDM contribution will lead
to a very weak perturbation to the light elements.

Combining the results of eqs.\ \pref{eq:Saha}, \pref{eqn:dDrate},
and \pref{eq:d-dest}
and requiring $\dot{Y}_d = 0$ gives the QSE deuterium abundance.
The deuterium bottleneck ends, and the light
elements are formed, when this increases to order unity, at the 
temperature
\beq
T_9 \approx \frac{25.81}{34.18 - \ln{(\eta_{10})} -
\frac{3}{2}\ln{(T_9)} + \EE{2.1}{-9}\eta_{10}\alpha_1\! \left(\frac{s}{\rm cm^2\ 
g^{-1}}\right)\!\left( \frac{\Omega_{\rm
D}}{\Omega_{\rm B}}\right) T_9}
\eeq
with $T_9 \la 1$.

Thus one can see that with $\frac{\Omega_{\rm
D}}{\Omega_{\rm B}} \sim 10$, one needs an
extremely large cross section, $\alpha_1 s \ga 10^{7}$ cm$^2$
g$^{-1}$, to
cause a noticeable change in the light element predictions. 
\he4 is the most sensitive to these changes.
To derive a limit on SIDM, we will adopt 
a very generous observational lower bound $Y_p \ge 0.230$
\cite{he_obs}.  
Using this, and a full numerical implementation of
eq.\ \pref{eqn:dDrate} in the expanded BBN code discussed in
\cite{cfo1}, yields a limit  
$\alpha_1 s \le \EE{1.263}{8} \ {\rm cm^2 \, g^{-1}}$
for the case $\md  \gg m_p$.
Given
the rate dependence on dark matter mass
and cosmological
parameters we can derive an explicit form for the maximum cross
section,
\beq
\alpha_1 s < \EE{1.263}{9}\frac{\ \left( 1 + \frac{m_d}{M_{\rm
D}}\right)^{3/2}}{\left( 1 + \frac{m_p}{\md }\right)^2\ } \left(
\frac{\Omega_{\rm B}}{\Omega_{\rm D}} \right) \ {\rm cm^2 \ g^{-1}}
\eeq
for a baryon-to-photon ratio $\eta_{10} = 5.8$ and
requiring $Y_p > 0.230$.  The constraints are only slightly different
for different assumed $\eta$ and helium abundances.

We see, therefore, that 
BBN cannot
place strong constraints on dark matter mass or strong baryon cross
section.
In other words, energy-independent SIDM is completely compatible with
light element constraints and BBN.
The weakness of the constraint follows from the
smallness of $\eta$ and its imposition of the deuterium bottleneck
which delays nuclear reactions until temperatures far below
the binding energies of the light nuclei.
This ensures that SIDM-deuteron interactions will be too weak
to allow dissociation, and the lack of SIDM-nucleon bound states
ensures that SIDM particles will not otherwise compete with
nucleons as the light elements are built up.

Having seen that SIDM-nucleon interactions
are compatible with nucleosynthesis in the early universe,
we now turn to a present-day consequence of these interactions,
namely the production of $\gamma$-rays from SIDM collisions
with cosmic rays.

%
%

\section{Limiting SIDM with Cosmic Rays}
\label{sect:CR}

The Galaxy is penetrated by a flux of nonthermal,
relativistic nuclei, the cosmic rays.
The bulk of these particles (i.e., those with $E \la 100$ TeV)
are of Galactic origin, and have kinetic 
energies typically $\sim 1$ GeV with 
a power law energy spectrum 
\begin{equation}\label{spectrum}
\frac{dN}{dE} \propto E^{-(\gamma+1)}
\end{equation}
where 
$\gamma \sim 1.7$.
Inelastic collisions of cosmic ray nuclei (primarily 
protons) with SIDM particles
are detectable in principle by the
gamma-ray signature of the
decay of the neutral pions produced in such interactions:
\[
\label{eq:pion-rxn}
\begin{array}{lrlll}
D + p \rightarrow &  D \,+& \!\!\! \Delta^{+}&  \\
 & & \!\!\hookrightarrow& p\, +&\!\!\!\pi^0 \\
 & & & & \hookrightarrow  \gamma + \gamma
\end{array}
\]
where $D$ is the dark matter particle. 
This reaction is simply the SIDM analog of
the usual pion production $pp \rightarrow pp\pi^0 \rightarrow \gamma\gamma$
reaction, responsible for the bulk of the
Galactic contribution to the $\gamma$-ray sky
\cite{stecker,stecker2,hunter,strong}.

The Galactic gamma-ray signature for this process
is readily calculated.  
We
assume axial symmetry for the cosmic-ray and SIDM spatial distributions,
and a relevant dark matter--nucleon cross 
section $\sigma_{DN}^{\rm inelastic}$ independent of the incoming
proton energy.  The 
emissivity $q_\gamma$
($\gamma-$ray photons per volume per time) is then
\begin{equation}\label{emis1}
q_\gamma(r,z)=2\,n_{_D}(r,z)\, \sigma_{DN}^{\rm inelastic}
 \, \Phi_p (r,z) \,\,.
\end{equation}
Here, the factor of 2  is the number of photons per $\pi^0$ decay, 
$n_{_D}(r,z)$ is the 
number density distribution of the dark matter in the Galaxy, 
and 
$\Phi_p$ is the angle-integrated cosmic ray proton 
intensity for all energies above the pion production threshold, 
\begin{equation} \label{protonflux}
\Phi_p=4\pi \int_{E_{\rm min}}^\infty \phi_{E,p} \,dE
\end{equation}
with 
\begin{equation}\label{thres}
E_{\rm min} = \left( m_p+m_{\pi^0}\right) \left[ 1 + 
\frac{m_{\pi^0} (2m_p +m_{\pi^0} )}{2\md  (m_p +m_{\pi^0})} \right]
\end{equation}
For $\md $ larger than $\sim 1$ GeV, the second term in the brackets of 
eq. (\ref{thres}) is $\ll 1$ and 
$E_{\rm min}$ does not change appreciably with $\md $, and
we have $E_{\rm min} \sim m_p+m_{\pi^0}$.
Using the parameter $s=\sigma_{DN}^{\rm elastic}/\md $, and
the fact that with typical dark matter halo speeds $v/c \sim 10^{-3} \ll 1$,
then $\rho_{D} = \md n_{D} $ and we 
can re-write eq. (\ref{emis1}) as
\begin{equation} \label{emis2}
q_\gamma(r,z)=2\,\rho_{_D}(r,z)\,\alpha_2 \, s \, \Phi_p (r,z) \,\,.
\end{equation}
The parameter $\alpha_2$ as used in eq. (\ref{emis2}), 
is the product of the ratio
$\sigma_{Dp \rightarrow DN\pi }/ \sigma{_{DN}^{\rm elastic}}$ 
and the
branching fraction $f_0$ for the decay of the $\Delta ^+$ to 
$p+\pi^0$ rather than $n+\pi^+$.
{}From isospin considerations it follows that
$f_0 = 2/3$.  In the absence of a detailed particle physics model for
the interacting dark matter, it is natural to expect  that the CR
inelastic collision strength is related simply to the breakup
collision strength by just this isospin factor, $\alpha_2 = 2/3 \
\alpha_1$, as both reactions are simply exciting internal degrees of freedom
in the proton. 

%

The resulting $\gamma-$ray intensity $\phi_{\gamma, _{DN}}$ in any given 
direction will then be the line integral of $q_\gamma$ along the
line of sight: 
\begin{equation} \label{int1}
\phi_{\gamma, _{DN}}=
\frac{1}{4\pi}
\int_{\rm l.o.s.} q_\gamma(r,z) \ {\rm d}\vec{\ell} 
\end{equation}
which, along a line of sight defined by a set of
Galactic coordinates ($\ell,b$), 
takes the form 
\begin{equation} \label{int2}
\phi_{\gamma, _{DN}}=
\frac{1}{4\pi}
\int_0^{\infty} 
q_{\gamma}\left(\sqrt{(R\cos b\cos \ell-L)^2+(R \cos b \sin \ell)^2 +
(R \sin b)^2}, R \sin b \right) \,dR
\end{equation}
where $L$ is the distance of the Sun from the Galactic center,
and $R$ is the heliocentric distance of a
point of Galactocentric distance $\vec{r}$.

A first estimate of the expected $\gamma$-ray intensity from cosmic ray - dark 
matter interactions can be made if we  
approximate the spatial distribution of the CR flux by a step-function in 
space: Uniform and equal to the demodulated solar value 
inside an oblate ellipsoid with a semi-major
axis 
of $10$ kpc in the Galactic plane and 1 (5) kpc towards
the Galactic poles,
and zero outside this ellipsoid. In the limit $\md  \rightarrow \infty$, 
eqs. (\ref{protonflux}) and (\ref{thres}) give 
$ \Phi _{p, \infty} = 11.8 \
{\rm cm^{-2} \,s^{-1}}$, while if we relax our assumption of large
$\md $,  $\Phi_p$ will scale as $\Phi_p = \Phi_{p,\infty} \left[ 1 + 
\frac{m_{\pi^0} (2m_p +m_{\pi^0} )}{2\md  (m_p +m_{\pi^0})}
\right]^{-\gamma}$

In addition, we assume that the mass distribution of the 
dark matter halo can be described as a flattened, non-singular isothermal
sphere \cite{olling}, 
$\rho(r,z)=\rho_c r_h^2 /\left(r_h^2+r^2+(z/q)^2\right)$ ,  
where $\rho_c$ is the central density, $r_h$ the core radius and
$q$ the flattening of the halo. In the following calculations 
we have used $q = 0.8$,
$\rho_c = 0.05 \, {\rm M_\odot \, pc^{-3}}$
and a value of $r_h = 5 \,{\rm kpc}$, such that the total dark 
matter mass included in a radius of $100 \, {\rm kpc}$ is
$\sim 10^{12} \, {\rm M_\odot}$. The distance of the Sun from the 
Galactic center was taken to be $L \approx 8.5 \,{\rm kpc}$.

We have calculated $\phi_\gamma$ in this simple approximation for 
directions of observation towards the Galactic center $(\ell=0^\circ, 
\,b=0^\circ)$, 
towards the anti-center $(\ell=180^\circ, \,b=0^\circ)$ and 
perpendicular to the Galactic plane $(b=0^\circ)$ and 
we have compared the results with the corresponding EGRET  
observational results from Hunter {\it et al} \cite{hunter} and  
Sreekumar {\it et al} \cite{sreekumar} to derive upper limits for
$\alpha_2 s$. Our results for the limit of large $\md $ are shown
in 
Table 1. The general result for the cross section upper limit will
be given by
\begin{equation}
\alpha_2 s < 
\alpha_2 s_{\infty} \left[
 1 + 
\frac{m_{\pi^0} (2m_p +m_{\pi^0} )}{2\md  (m_p +m_{\pi^0})} 
\right]^\gamma
\end{equation}
where $s_{\infty}$ is the value
as $\md \rightarrow \infty$.

\begin{table}
\begin{tabular}{||l||c|c|c||} \hline\hline
Direction of & SIDM expectation & EGRET observation & implied
$\alpha_2 s$\\
Observation & 
$\left( \frac{\alpha_2 s}{\rm 1 cm^2 /g  }\right) \frac{\rm photons}{\rm cm^2
\,\,sr\,\,s}$
		& $\frac{\rm photons}{\rm cm^2 \,sr\,s}$
					           & upper limit \\\hline
\hline
Galactic center & $2.1 \times 10^{-1}$ & $5.5 \times
10^{-4}$ &
			$2.6 \times 10^{-3}$ \\
\hline
Galactic anti-center & $6.7 \times 10^{-3}$ & $1.5
\times 10^{-4}$
			& $2.2 \times 10^{-2}$  \\
\hline
Galactic poles & 
$2.5(12.) \times 10^{-3}$  
			    & $1.3 \times 10^{-5}$ & $5.2 (1.1) \times 10^{-3}$ \\
\hline
M31 & $9 \times 10^{-2}$ & $< 8 \times 10^{-5}$ & $9\times10^{-4}$ \\
\hline \hline
\end{tabular}

\caption{Expected $\gamma$-ray intensities from SIDM-CR interactions and 
implied cross-section upper limits. 
The numbers in the parentheses indicate results for a 
semi-major axis of the cosmic ray halo ellipsoid in the 
direction perpendicular to the Galactic plane equal to 5
kpc, while other results refer to a semi-major axis of 1 kpc. }

\end{table}

Note that for a sufficiently flattened cosmic ray halo 
(1 kpc minor axis) the strongest 
constraint comes not from the direction where the $\gamma-$ray 
intensity is 
minimum (the Galactic poles), but from the observations towards
the Galactic center, where the $\gamma-$ray intensity is maximum.

For the derivation of the upper limits presented above, 
the entire $\gamma-$ray 
flux detected by EGRET was attributed to pion decay from 
cosmic ray--SIDM interactions. 
This is a good assumption 
for the direction towards the Galactic poles where the 
contribution of the $\gamma-$rays from collisions of cosmic rays
and interstellar baryons is not dominant.
However, in the Galactic plane most of the observed intensity
can be attributed to the presence of interstellar baryonic gas. Thus, the
constraints in the directions of the Galactic center and anti-center 
could be made
even more stringent than these conservative limits
if one introduces a model for the emission from
the interstellar medium.

We note that constraints of this type can be derived
for external galaxies as well.  In particular, we consider
M31, which has the advantages of being nearby, and
similar to the Milky Way. 
M31 has been searched for in {\it EGRET} sky maps, 
but no positive detection has been possible and an upper limit 
of $1.6 \times 10^{-8} \ {\rm cm^{-2} \ s^{-1}}$ has been
placed instead on its diffuse $\gamma-$ray flux \cite{blom}.
Given the measured dark mass $M_D^{\rm M31}  = 6 \times 10^{10} \msol$ in 
the inner 10 kpc of M31 \cite{m31},
we predict a SIDM $\gamma$-ray flux of
\beq
\phi_{\gamma}^{\rm M31} 
  = 2 \frac{\Phi_p^{\rm M31} \alpha_2 s M_D^{\rm M31} }{4\pi R_{\rm M31}^2}
\eeq
where $R_{\rm M31} = 750$ kpc is the distance to M31.
Following \cite{pf}, we estimate the M31 cosmic ray flux
$\Phi_{p}^{\rm M31}$ by assuming supernovae to be
the site of cosmic ray acceleration,
and thus taking the cosmic ray flux to be proportional
to the supernova rate.  This gives
$\Phi_{p}^{\rm M31} = 0.45 \Phi_{p}^{\rm MW}$,
which allows us to predict
$\Phi_{\gamma}^{\rm M31} = 1.8 \times 10^{-5} \ \alpha_2 s \ {\rm cm^{-2}
\
s^{-1}}$ (where s is given in ${\rm cm^2 / g}$).
This must be lower than the observed limit, 
which implies a limit
\beq
\alpha_2 s ^{\rm M31} \la 9 \times 10^{-4} {\rm cm^2 \ g^{-1}}
\eeq
Thus we see that the M31 $\gamma$-ray limits are about an order of
magnitude more constraining than those of the Milky Way;
however, since they rely on more assumptions
regarding the cosmic ray flux in M31, 
we regard the Milky Way limits as more secure.

Finally, it is very intriguing to note that
a recent reanalysis \cite{dixon} of the {\em EGRET} data
finds evidence for a diffuse Galactic $\gamma$-ray halo.
Specifically, a wavelet analysis was used to
identify a $\gamma$-ray component 
at large angular scales, centered on the Galactic center,
the intensity of which significantly exceeds the
predictions of a model which includes known Galactic sources.
One should bear in mind the error budget is large:
the inferred halo intensity levels are uncertain to within
a factor of $\sim 2$, and possible systematic effects remain.
Also, it is entirely possible that the 
halo can be explained via inverse Compton scattering of
background photons off of energetic cosmic ray electrons
\cite{strong}.

Nevertheless, it is tantalizing to interpret this diffuse halo
in terms of SIDM-baryon interactions.
The $100$ MeV halo intensity levels toward the poles is
quite uncertain but of order
$\sim 10^{-6} \ {\rm cm^{-2} \ s^{-1} \ sr^{-1}}$,
i.e., about 10\% of the full {\em EGRET} result.
In terms of our model, this would amount to
a {\em measurement} of $\alpha_2 s$ at about
10\% of our limit in Table 1, namely
$\alpha_2 s \sim 5(1) \times 10^{-4} \ {\rm cm^2 \ g^{-1}}$.
Future $\gamma$-ray observatories such as GLAST
can verify the existence of this diffuse Galactic halo,
and will be in a position to measure its spectrum.
If indeed the flux arises from $\pi^0$ production in 
SIDM-baryon interactions, we would predict that the 
spectrum should reflect this origin, in particular,
it should be a smooth continuum which shows the ``pion bump'' feature at
$E_\gamma = m_{\pi^0}/2 = 67.5$ MeV.
In addition, similar halo emission should be
visible around Local Group galaxies--not only M31, but
also the Magellanic Clouds.

%
%

\section{Other Possible Constraints}
\label{sect:other}

SIDM-nucleon interactions can have other effects as well.
The possibility of DM-baryon interactions contributing to the gamma ray
background provokes the question whether some fraction of the $\sim 20$
unidentified  EGRET sources at high
Galactic latitudes could be due to dark matter concentrations in the Milky
Way halo. A rough estimate shows that typical gamma-ray fluxes of these
sources would be compatible with DM clumps of order 
$10^3 M_{\odot} \ (R/1 {\rm \, kpc}) \ (1 { \rm \, cm^2 \,
g^{-1}} /\alpha_2s)$.
While this is an intriguing
speculation, the small 
number statistics of these sources makes distinguishing possible
candidates from a 
background of approximately isotropically distributed AGN a difficult
task. A more detailed study will therefore have to await the higher
sensitivity and resolution data from future missions such
as GLAST.

Another potential (but model-dependent) constraint comes 
from diffuse radiation. Consider the effect of SIDM particle passage 
through the interstellar medium of our Galaxy, most of
which is hydrogen.
Elastic collisions between SIDM particles and the protons 
occur, with  
a relative velocity given by 
$v \sim 200 \ {\rm km/s} \sim 10^{-3} c$, and
thus a center-of-mass energy $E_{\rm CM} \simeq m_p v^2/2 = 0.2$ keV.
Given that we expect $m_p \ll \md $, it follows that
a sizable fraction of this energy will be imparted to the
recoiling proton.  As a source of heating, 
this process does not violate observed constraints
on the properties of molecular clouds \cite{sged}.
However, if the scattering process leads
to photon emission by the proton, this radiation
might be observable.
One can show that if each scattering event
produces a photon with $E_{\rm CM}$,
then the resulting diffuse emission is at or above the level
observed in soft X-rays.  Of course, it need not be
the case that this radiative scattering occurs, but
given a particular SIDM model, this constraint should
be investigated.

\begin{figure}
\epsfig{file=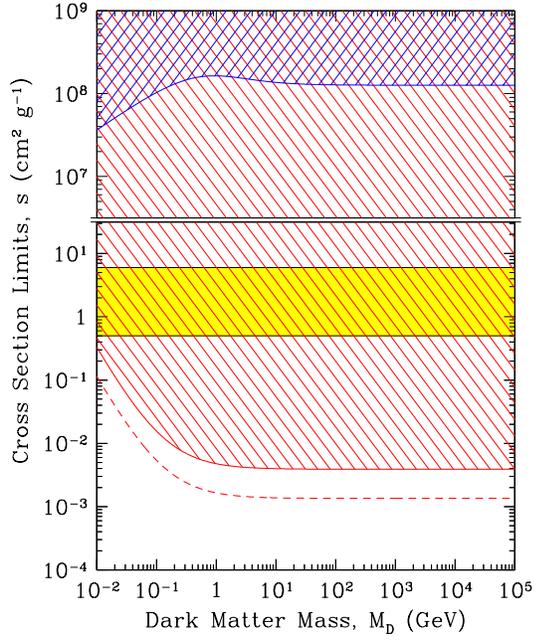,height=4in,angle=0}
\caption{Limits on the cross section-to-mass ratio
$s$ as a function of SIDM mass.
The values needed for galaxy halos 
(eq.\ \ref{eq:s-halo}) are shown
in the shaded region.
Excluded regions are hatched; single-hatched
regions 
are excluded due to the inelastic interactions with
Galactic cosmic rays; the cross-hatched regions are
also excluded via BBN.
The dashed curve shows the limit using M31 
$\gamma$-ray observations.
\label{fig:s-m}}
\end{figure}

%
%

\section{Discussion and Conclusions}
\label{sect:dis}

\begin{figure}
\epsfig{file=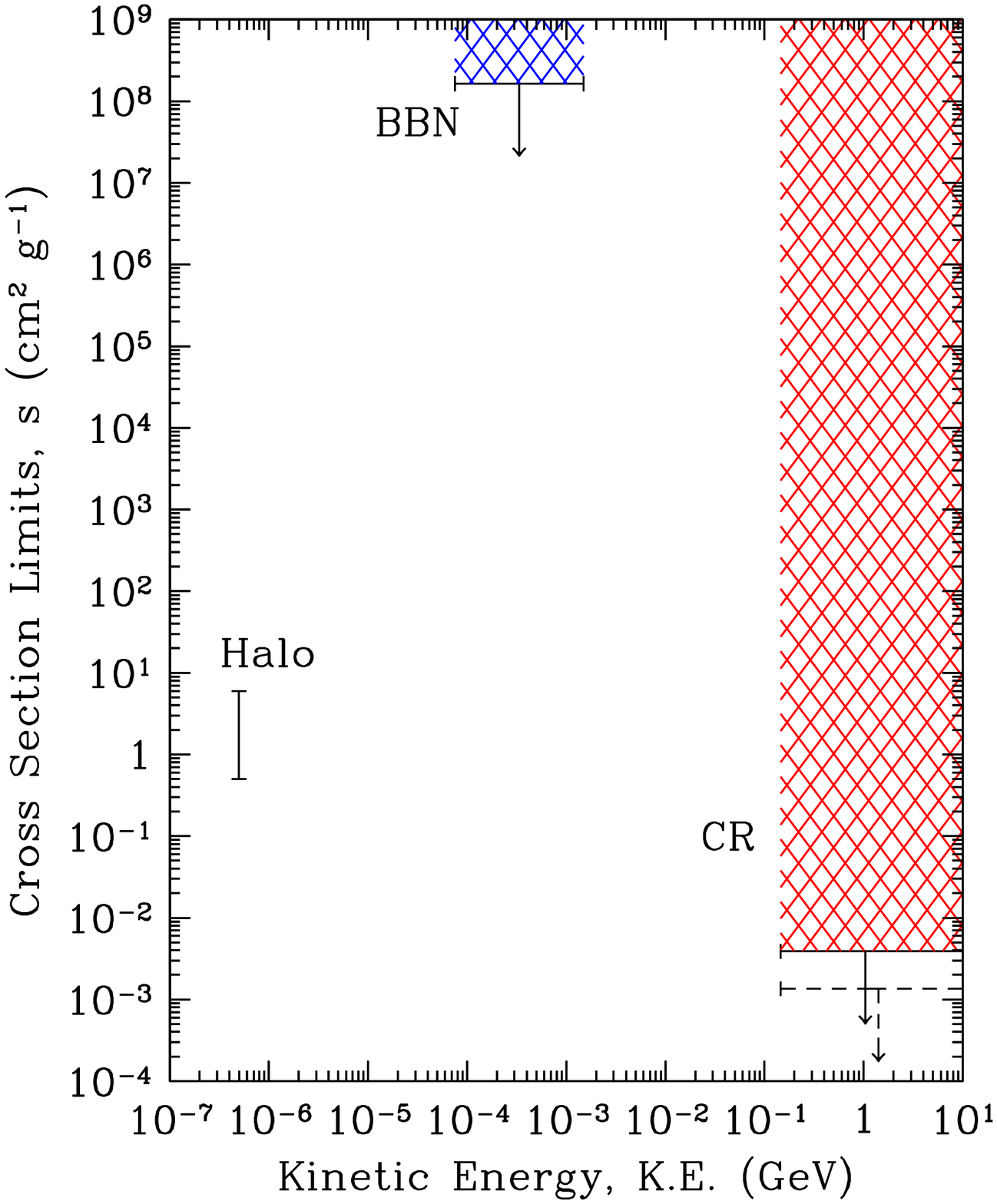,height=4in,angle=0}
\caption{Limits on the cross section-to-mass ratio
$s$ as a function of center of mass energy,
for the case in which $\md  \gg m_p$.  The dashed curve shows the
limit using M31 $\gamma$-ray observations.
\label{fig:s-E}}
\end{figure}

We have considered constraints on strong interactions
between dark matter particles and baryons. 
The big bang nucleosynthesis (BBN) constraints are weak, indicating that
interacting dark matter, as it has been recently proposed, is fully 
consistent with BBN.  However, very strong limits
come from considering cosmic ray (CR) interactions with the dark matter.
For dark matter masses $\md  \ga 1$ GeV the ratio $s\equiv
\sigma_{DN}/\md $ is limited to be less than $3.9\times 10^{-3}$cm$^2/$g.
Constraints from direct detection experiments such as the quantum
calorimeter XQC \cite{mcc96} require
$\md \ga 10^5$ GeV\cite{wand,mcg}. 
Our limit therefore constrains dark-matter baryon interactions to be
less than a hundredth
of the self-interaction strength which, simulations suggest, 
is required to affect the 
structure of galaxy halos within the self-interacting dark matter (SIDM)
scenario.  Any particle physics model of SIDM must explain
this order of magnitude suppression of the dark matter-baryon interaction.

For the case in which the inelastic cross section is
independent of energy, our results are summarized
in Figure \ref{fig:s-m}.
It is of course possible, and indeed plausible, that
SIDM-baryon interactions might be energy-dependent.
Thus, we have displayed in
Figure \ref{fig:s-E}
the constraints on $s$ as a function of
the center-of-mass energy at which those constraints
apply.  Both figures show that energy-independent
DM-baryon interactions are ruled out by the cosmic ray
constraints, unless the ratio of DM-baryon scattering to DM
self-scattering $\alpha \la 10^{-3}$.  On the other hand,
if the interaction is energy-dependent,
Figure \ref{fig:s-E} constrains the behavior.
For example, if we assume that $s \propto E_{\rm CM}^{n}$,
where $E_{CM}$
is the center of mass kinetic energy,
then we find using our limits from cosmic rays coming from the
Galactic center that $n < -0.39$.  Using the M31 constraint
steepens this to $n< -0.47$.  In both cases,
a $1/v$ scaling (or stronger dropoff) is 
allowed.   Thus,  our $\gamma$-ray constraints provide important information
about the energy dependence of a putative SIDM-baryon interaction.
If such an energy dependent cross-section were characteristic of dark matter
self-scattering as well,  SIDM would be less interactive in
objects with high velocity dispersion such as clusters of
galaxies, possibly softening constraints on SIDM from the statistics of
strongly lensed arcs \cite{arcs}.

We stress that our results do not rely on any detailed assumptions about
the underlying particle physics model governing the interactions. 
In the context of a more detailed particle physics model for SIDM,
one could also examine signatures of less generic processes, for example
the generation of diffuse emission of soft X-rays from  
radiative scattering between SIDM and interstellar medium particles.

Finally, we note that elastic SIDM-baryon interactions  thermalize the
dark matter at early times. This may have
potentially important consequences for structure formation.
We plan to investigate this issue in more detail in a future publication.

\vskip 0.5in
\vbox{
\noindent{ {\bf Acknowledgments} } \\
\noindent  
We thank 
Paul Steinhardt for his encouragement and insight, and
Jim Buckley for helpful conversations on
gamma-ray observations.
The work of V.P.  was partially supported by a scholarship
from the Greek State Scholarship Foundation.
The work of B.D.F., V.P., and R.H.C.
was supported by National Science Foundation
grant AST-0092939.
}

\end{document}